\documentclass{article}
\usepackage{amssymb}
\newcommand{\be}{\begin{equation}}
\newcommand{\ee}{\end{equation}}
\newcommand{\bc}{\begin{center}}
\newcommand{\ec}{\end{center}}
\newcommand{\ber}{\begin{eqnarray}}
\newcommand{\ear}{\end{eqnarray}}
\newcommand{\ba}{\begin{array}}
\newcommand{\ea}{\end{array}}

\newcommand{\al}{\alpha}

\newcommand{\bx}{\Box}

\newcommand{\de}{\delta}

\newcommand{\ep}{\epsilon}

\newcommand{\fr}{\frac}
\newcommand{\Ga}{\Gamma}

\newcommand{\hb}{\hbar}

\newcommand{\La}{\Lambda}
\newcommand{\la}{\lambda}
\newcommand{\lb}{\label}

\newcommand{\Lg}{{\cal L}}

\newcommand{\n}{\nonumber\\}

\newcommand{\na}{\nabla}
\newcommand{\Or}{{\cal O}}

\newcommand{\om}{\omega}

\newcommand{\ps}{\psi}

\newcommand{\Si}{\Sigma}
\newcommand{\si}{\sigma}
\newcommand{\sq}{\sqrt}
\newcommand{\st}{\stackrel}
\newcommand{\ta}{\tau}

\newcommand{\te}{\theta}

\begin{document}
\title{Quantizing the Line Element Field.}
\author{Mark D. Roberts,
54  Grantley Avenue,  Wonersh Park GU5 OQN,\\
            {mdrobertsza@yahoo.co.uk}, 
{http://cosmology.mth.uct.ac.za/$\sim$~roberts}}
\maketitle
\bc Eprint:  gr-qc/0302121  \ec
\bc Comments:  9 pages.\ec
\begin{abstract}
A metric with signature (-+++) can be constructed from a metric with signature (++++)
and a double-sided vector field called the line element field.   
Some of the classical and quantum properties of this vector field are studied.
\end{abstract}
{\small\tableofcontents}
\section{Introduction.}\lb{intro}
The difference between a Lorentzian and a positive definite metric can be 
expressed as a double-sided vector field $U$ called the line element field.
This is done in \cite{HE}p.38,  
but there only the ratio of vectors in the two spaces is considered;
this appears to be the only reference on the line element field 
and it is from it that the nomenclature is taken.
Thus the study of fields or extented objects in Lorentzian spacetime is reduced to the study
of the same object in a positive definite space 
and the study of the corresponding line element field. 
In particlular this can be done for gravity,   
where the positive defininte action is sometimes called the Euclidean action \cite{GH}.
Things not looked at here include:
{\it firstly} any relationship to analytic continuation,
whether for quantum field theory on curved spacetime 
or for the energy condition \cite{HE}p.89,
{\it secondly} any classical or quantum detailed mechanism or perturbation whereby 
a positive definite space could change to a Lorentzian spacetime,   
for example in the early universe,
{\it thirdly} the connection with the Kubo-Martin-Schwinger \cite{emch} condition
where the transformation $\ta\rightarrow it$ has thermal properties,
{\it fourtly} a quantized line element field might fluctuate,
this fluctuation could be thought of in terms of the tetrad 
rather than the metric,  leading to fluctuating null cones,
compare Penrose \cite{penrose},
{\it fifthly} any comparison with 
the Toll \cite{toll} - Scharnhorst \cite{scharnhorst} effect,  
where fluctuations in the quantum electrodynamical vacuum 
cause fluctuations in the speed of light,
{\it sixthly} any comparison with the average size of these fluctuations,
compare Ellis {\it et al} \cite{EMN} and Yu and Ford \cite{YF},
{\it seventhly} not only can the difference between the two signatures 
be thought of as a vector field,  also the difference between tensors constructed
from the resulting metrics is tensorial,   the Bianchi identities will also differ
by a tensorial object constructed from the line element field and this gives another
way of investigating conservation laws for the two signatures,  compare \cite{HD}.

In \S\ref{curv} some examples of positive definite metrics are presented and how to change
their signature via a vector field is shown;  
this is successively generalized to vanishing shift metrics 
and then the general theory,
next the first derivatives are studied, 
and expressed in terms of a contorsion tensor.
The second derivatives of the metric are governed by
the Riemann tensor which can be expressed by independent terms in the contorsion
and Christoffel contection.
In \S\ref{quant} the Einstein-Hilbert action is decomposed 
into a postitive definite part and a line element field part,
the line element field part is varied with repect to both $U$ and $\dot{U}$.
The variation with respect to $\dot{U}$ gives the momentum.
Quantization is implemented by replacing this momentum by a differential operator
to give a modified Klein-Gordon equation.
Then the lowest order approximation to the modified Klein-Gordon equation is calculated,
and the wavefunction is calculated for some specific spacetimes.

Notation used includes the bracket notation of \cite{HE}p.20 
\be
2V_{(a,c)}=V_{a,c}+V_{c,a}=2V_{(a;c)}+2\{^e_{ca}\}V_e,~~~~~~~~
2V_{[a,c]}=2V_{[a;c]}=V_{a,c}-V_{c,a},
\lb{i1}
\ee
the scalars constructed from the expansion and vorticity 
\be
\te\equiv\te^a_{.a},~~~~~
\om^2\equiv\om^{ab}\om_{ab},~~~~~
\si^2\equiv\si^{ab}\si_{ab},
\lb{i2}
\ee
and vector fields 
\ber
&U_a ~~~{\rm for~ a~ general~ vector},\n
&V_a ~~~{\rm for~ a~ normalization~ of~ this~ to~} \pm1,\n
&W_a ~~~{\rm for~ a~ specific~ vector}.
\lb{i3}
\ear
\section{Curvature.}\lb{curv}
For a given positive definite metric $\st{-+}{p}_{ab}$ and vector field $\st{++}{U}_a$,
one can construct a Lorentzian spacetime with covariant metric
\be
\st{-+}{g}_{ab}=-2\fr{\st{++}{U}_a\st{++}{U}_b}{\st{++}{U}^2}+\st{++}{p}_{ab},
\lb{n1}
\ee
This can be illustrated using the positive definite Schwarzschild metric
\be
\st{++}{ds}^2=+\left(1-\fr{2m}{r}\right)dx^2_0+\left(1-\fr{2m}{r}\right)^{-1}+r^2d\Si^2_2,~~~
\st{++}{W}_a=\sq{1-\fr{2m}{r}},~~~
\st{++}{W}^a\st{++}{W}_a=+1,
\lb{pdsc}
\ee
or the positive definite Robertson-Walker metric
\be
\st{++}{ds}^2=+dx^2_0+R^2d\Si^2_3,~~~
\st{++}{W}_a=(1,0),
\lb{pdrw}
\ee
then using \ref{n1} the spacetime metric is recovered. 
Instead of $U$ it is often convenient to work with the unit vector
\be
V_a=\fr{U_a}{\sq{\pm U^2}},~~~
U^2=U^aU_a,
\lb{n4}
\ee
There is a problem of what the contravariant form of $\st{-+}{g},~p ~\& ~U$ should be.
Say we are given a positive definite space with shift-free metric and vector field
\ber
&\st{++}{p}_{ab}=(p^2,p_{ij}),~~~~~
\st{++}{p}^{ab}=\left(\fr{1}{p^2},p^{ij}\right),~~~~~
\det(\st{++}{p}_{ab})=p\det(p_{ij}),\n
&\st{++}{W}_a=(p,0),~~~~~
\st{++}{W}^a=\left(\fr{1}{p},0\right),
\lb{eq1}
\ear
where for simplicity there are no cross terms $(\ta,i)$ terms,
as that would require either $W_a$ or $W^a$ to be no longer one component.
Now one can construct a Lorentzian spacetime with covariant metric
\be
\st{-+}{g}_{ab}=(-p^2,p_{ij}),~~~~~
\det(\st{-+}{g}_{ab})=-\det(\st{++}{p}_{ab}).
\lb{eq2}
\ee
Consistency seems to require
\be
\st{-+}{g}^{ab}=-2\st{++}{V}^a\st{++}{V}^b+\st{++}{p}^{ab}
=\left(-\fr{1}{p^2},p^{ij}\right),
\lb{eq3}
\ee
note that only cross terms in $V$ occur so perhaps $\st{-+}{V}$ could have been used,
\ref{c5} shows that this is not the case.
Taking $\st{-+}{V}_a=\st{++}{V}_a$ and raising using this metric
\be
\st{-+}{W}^a=\left(-\fr{1}{p},0\right),~~~~~
\st{-+}{W}^2=-1,
\lb{3b}
\ee
so that $\st{-+}{V}$ is a timelike vector.
Similiarly taking $\st{-+}{p}_{ab}=\st{++}{p}_{ab}$ 
and raising indices using \ref{eq3} gives
\be
\st{-+}{p}^{ab}=\st{++}{p}^{ab}.
\lb{3c}
\ee
Some products using the above tensors are
\ber
&\st{-+}{g}^c_{~c}\equiv\st{-+}{g}^{ab}\st{-+}{g}_{bc}
=\st{++}{p}^a_{~c}-2\st{++}{V}^a\st{++}V_c,~~~
\st{++}{p}^{ab}\st{-+}{g}_{bc}=2\st{-+}{V}^a\st{-+}{V}_c+\st{-+}{g}^a_{~c},\\
&-\st{-+}{V}_a=\st{-+}{V}^b\st{-+}{g}_{bc},~~~
\st{-+}{V}^a=\st{-+}{V}^c\st{-+}{g}^a_{~c}
=-\st{-+}{V}_c\st{++}{p}^{ab},~~~
\st{++}{p}^{ac}\st{-+}{V}_c=-\st{-+}{V}^a.\nonumber
\lb{3e}
\ear
Collecting this together consistency requires
\be
\st{-+}{g}^{ab}=-2\fr{\st{++}{U}^a\st{++}{U}^b}{\st{++}{U}^2}+\st{++}{p}^{ab},
\lb{n2}
\ee
and
\be
\st{-+}{V}_a=\st{++}{V}_a,~~~
\st{-+}{V}^a=-\st{++}{V}^a,~~~
\st{-+}{p}_{ab}=\st{++}{p}_{ab},~~~
\st{-+}{p}^{ab}=+\st{++}{p}_{ab}.
\lb{n3}
\ee

The above system is new and slips can be made by 
relying on ones intuition from studying spacetimes using the projection tensor,  
see \ref{c7} below,
or confusing the (-+) and (++) spaces.
The most common of these is
\be
4=g^a_a\st{?}{=}-2V_aV^a+p^a_a=-2V_aV^a+4,
\lb{c4}
\ee
suggesting that $V$ is null, contrary to assumption.
The correct calculation is
\be
4=\st{-+}{g}^{ab}\st{-+}{g}_{ab}
=\left(-2\fr{\st{++}{U}^a\st{++}{U}^b}{\st{++}{U}^2}+p^{ab}\right)
 \left(-2\fr{\st{++}{U}_a\st{++}{U}_b}{\st{++}{U}^2}+p_{ab}\right)
=\left(4\fr{\st{++}{U}^2}{\st{++}{U}^2}-2-2\right)\fr{\st{++}{U}^2}{\st{++}{U}^2}+4,
\lb{c5}
\ee
which also serves as showing that $\st{++}{U}$ rather than $\st{-+}{U}$ should be used
for $\st{-+}{g}$ in \ref{n2}.
The projection tensor is defined as
\be
\st{-+}{h}_{ab}\equiv\st{-+}{g}_{ab}+\fr{\st{-+}{U}_a\st{-+}{U}_b}{-\st{-+}{U}^2}
=\st{++}{p}_{ab}-2\fr{\st{++}{U}_a\st{++}{U}_b}{\st{++}{U}^2}-
                  \fr{\st{-+}{U}_a\st{-+}{U}_b}{\st{-+}{U}^2}
=\st{-+}{p}_{ab}+\fr{\st{-+}{U}_a\st{-+}{U}_b}{\st{-+}{U}^2}
=\st{++}{p}_{ab}-\fr{\st{++}{U}_a\st{++}{U}_b}{\st{++}{U}^2},
\lb{c7}
\ee
where in this case the indices can be raised and lowered without change of form.

Having formed the metric the next problem is the properties of its first derivatives.
To form the connection $\Ga$ with $g$ in terms of the connection $\{\}$ with $p$ one has
\ber
&2\st{g}{\Ga}_{abc}\equiv g_{ba,c}+g_{ca,b}-g_{bc,a}\\
&=2\{_{abc}\}+4U^{-2}\left(-U_a\{^e_{bc}\}U_e
-U_aU_{(b;c)}+U_bU_{[c;a]}+U_cU_{[b;a]}\right)\n
&+4U^{-4}U_e\left(2U_aU_{(b}U^e_{.c)}-U_bU_cU^e_{.a}\right),\nonumber
\lb{eq4}
\ear
where the bracket notation \ref{i1} is used
and the covariant derivatives on the rhs of \ref{eq4} are formed with $p$.
Raising with the metric \ref{eq3}
\be
\Ga^a_{bc}\equiv\st{-+}{g}^{ad}\Ga_{dbc},
\ee
so that
\be
\Ga^a_{bc}=\{^a_{bc}\}+L^a_{~e}\{^e_{bc}\}+K^a_{bc},
\lb{eq6}
\ee
it is found that $L=0$ implying that the system is covariant.
From \ref{eq6} there is the relation between the covariant derivatives.
\be
\st{++}{V}_{a\st{-+}{;}b}=\st{++}{V}_{a\st{++}{;}b}-K^c_{ab}\st{++}{V}_c,
\lb{6b}
\ee
The contorsion tensor $K$ is
\ber
&2K^a_{bc}\equiv+8U^{-2}U_{(b}J_{c)}^{~~a}+4U^{-2}U^aU_{(bc)}-2U^{-4}M^a_{bc},\n
&J_{ab}\equiv U_{[a;b]}-2U^{-2}U_bU^cU_{[a;c]},\\
&M^a_{bc}\equiv(U^2)^aU_bU_c+(U^2)_bU^aU_c+(U^2)_cU^aU_b-2U^{-2}U^aU_bU_cU^e(U^2)_e,\nonumber
\lb{eq7}
\ear
there is no term in the accelerations $\dot{U}^a,~\dot{U}^b~{\rm or}~\dot{U}^c$,
$M$ vanishes if $U$ is a constant vector. 
Some properties of the contorsion $K$ are
\ber
&K^a_{ac}=0,~~~
K^a_{ba}=0,~~~
K^a_{bc}=K^a_{cb},~~~
U^bU^cK^a_{bc}=0,\n
&U_aK^a_{bc}=2U_{bc}-2U^{-2}(U_{(b}\dot{U}_{c)}+U_eU_{(b}U^e_{.c)})
                    +2U^{-4}U_bU_cU_e\dot{U}^e,
\lb{7e}
\ear
the dot being formed with the $p$ covariant derivatives;
such sysytems involving a connection and a contorsion occur repeatedly
in the study of curvature;  for instance in geometries involving torsion and/or
metricity such as the geometries of Weyl and Schouten \cite{schouten},
the study of a conformal factor,
and the study of weak metric perturbations \cite{mdr32}.
Alternatively the contorsion can be expressed in terms of the decomposed vector field
for a (-+) Lorentzian spacetime define $\te$ as in \ref{i2} and
\be
\om_{ab}\equiv h_a^{~c}h_b^{~d}V_{[c;d]},~~~
\te_{ab}\equiv h_a^{~c}h_b^{~d}V_{(c;d)},~~~
\si_{ab}\equiv\te_{ab}-\fr{1}{3}\te h_{ab},~~~
\dot{X}_{abc\dots}\equiv V^eX_{abc\dots;e},
\lb{eq8}
\ee
which allow the covariant derivative of a (-+) spacetime to be decomposed
\be
U_{a;b}=\te_{ab}+\om_{ab}+U^{-2}U_b\dot{U}_a+U^{-2}U_aU^eU_{eb}-U^{-4}U_aU_bU^e\dot{U}_e,
\lb{8b}
\ee
choosing a constant vector field this reduces to \cite{HE}eq.4.17.
Now the equations \ref{eq7} are in the (++) space and \ref{eq8} are in a (-+) spacetime;
they can be related using the projection tensor \ref{c7}.
The projections of the covariant derivative are
\ber
&h^c_ah^d_bU_{(c;d)}=U_{(a;b)}-U^{-2}U_{(a}\dot{U}_{b)}
                     -\fr{1}{2}U^{-2}U_{(a}(U^2)_{b)}+\fr{1}{2}U^{-4}U_aU_b\dot{(U^2)}\n
&h^c_ah^d_bU_{[c;d]}=U_{[a;b]}-U^{-2}\dot{U}_{[a}U_{b]}
                      -\fr{1}{2}U^{-2}U_{[a}(U^2)_{b]}
\lb{h8}
\ear
To transfer these quanties to the (++) space,  
the projection tensor \ref{c7} shows that 
it is only necessary to note that the negative quantity $U^2=\st{-+}{U}_a\st{-+}{U}^a<0$
is changed to the positive quantity $U^2=\st{++}{U}_a\st{++}{U}^a>0$,
and also that the covariant derivative in the expansion is changed using \ref{6b}.
$U_eK^e_{ab}$ is $-2$ times $\te_{ab}$ so that the sign of $\te_{ab}$ in the (++)
space is the negative of the form in \ref{h8} and from \ref{eq8}.
In particular
\be
-\te=U^a_{.;a}-\fr{\dot{(U^2)}}{U^2}.
\lb{ah8}
\ee
Using \ref {eq7} and \ref{h8} the contorsion tensor is found to be
\be
J_{ab}=\om_{ab}-\fr{\dot{U}_{(a}U_{b)}}{U^2}+\fr{U_{(a}(U^2)_{b)}}{2U^2},~~~
K^a_{bc}=\fr{4}{U^2}U_{(b}\om_{c)}^{~~a}-\fr{2}{U^2}U^a\te_{bc}
-2\sq{U^2}\dot{\left(\fr{U^a}{\sq{U^2}}\right)}\fr{U_bU_c}{U^4}.
\lb{n9}
\ee

The second derivatives are governed by 
the Riemann tensor which is, c.f.eq.111\cite{mdr32}
\be
\st{g~}{R^a}_{bcd}=\st{p~}{R^a}_{bcd}+2K^a_{.[d|b|;c]}+2K^a_{.[c|e|}K^e_{.d]b}.
\lb{eq10}
\ee
\section{Quantization.}\lb{quant}
Contracting the expression for the Reimann tensor \ref{eq10} 
and using \ref{eq2} the Einstein-Hilbert Lagrangian is
\be
\Lg_H=\sq{-g}\st{g}{R}=\sq{\det(p_{ab})}\left[\st{p}{R}
     +2K^{a~b}_{.[b|.|;a]}+2K^a_{.[a|e|}K^{eb]}_{..~~b}\right]
=\Lg_1+\Lg_2+\Lg_3.
\lb{eq11}
\ee
$\Lg_1$ is the positive definite action,  sometimes called the Euclidean action 
which has previously been studied,  $\Lg_2~\&~\Lg_3$ are new and 
easiest to describe in terms of the decomposed vector quantities \ref{eq8}
\ber
\Lg_2=-\fr{2}{U^2}(\dot{\te}-\te^2)
-2\left(\fr{1}{\sq{U^2}}\left(\fr{U^a}{\sq{U^2}}\right)^\circ\right)_a
\equiv l_1+l_2+l_3,\n
\Lg_3=\fr{4\om^2}{U^2}-\fr{4}{U^4}\left[\left(\fr{U^a}{\sq{U^2}}\right)^\circ U_a\right]^2
\equiv l_4+l_5,~~~
\om^2\equiv\om^{ab}\om_{ab},
\lb{eq12}
\ear
expanding $l_5=0$.
Varing with respect to $U$,
\ber
&\fr{\de l_1}{\de U^c}=
&-2U^{-2}\te_c+2\left(\left(U^{-2}U^eU^2\right)_e\right)_c
-2U^{-2}\left(U^{-2}U^e\right)_e(U^2)_c\n
&&-4U_c\left(U^{-2}\left(^{-2}U^f\right)_fU^e\right)_e
-4U^{-4}\left(U^{-2}U^e\right)_e(U^2)^\circ U_c,\n
&\fr{\de l_2}{\de U^c}=
&4\left(U^{-2}\te\right)_c
+4U^{-4}\te(U^2)_c-8U_c\left(U^{-4}\te U^e\right)_e
-8U^{-6}(U^2)^\circ U_c,\n
&\fr{\de l_4}{\de U^c}=
&-8\left(U^{-2}\om^{ce}\right)_e
+4U^{-2}\om^{ce}\left(2\dot{U}_e-U^2(U^2)_e\right),\n
&&\fr{\de l_3}{\de U^c}=\fr{\de l_5}{\de U^c}=0.
\lb{12b}
\ear
Varing with respect to $\dot{U}^c$
\be
\fr{\de l_1}{\de \dot{U}^c}=-\fr{4\te U_c}{U^4},~~~
\fr{\de l_2}{\de \dot{U}^c}=+8\fr{\te U_c}{U^4},~~~
\fr{\de l_3}{\de \dot{U}^c}=
\fr{\de l_3}{\de \dot{U}^c}=\fr{\de l_3}{\de \dot{U}^c}=0.
\lb{12c}
\ee
Therefore
\be
\Pi_a=\fr{\de}{\de \dot{U}^a}(l_1+l_2)=\fr{4\te U_a}{U^4}.
\lb{eqm}
\ee
This equation is not fully invertable $U_a=f(\Pi)\Pi_a$,
but is partially invertible $U_a=f(\Pi,U)\Pi_a$,
\be
\fr{U_a}{\sq{U^2}}=\fr{\Pi_a}{\sq{\Pi^2}},
\lb{eqp1}
\ee
and partially invertible $U_a=f(\Pi,\te)\Pi_a$,
\be
U_a=(4\te)^\fr{1}{3}\Pi^{-\fr{4}{3}}\Pi_a.
\lb{eqp2}
\ee
\ref{eqm} gives the constraint
\be
\la=\Pi_c\Pi^c-\fr{16\te^2}{U^6}.
\lb{eqco}
\ee
This is the only constraint so that quantization can be achieved via
\be
\Pi_a\rightarrow-i\hb\na_a
\lb{eqq}
\ee
with $U$ and hence $\te$ remaining unchanged.
Plancks constant $\hb$ is of the same dimensions as action,
explicitly $Mass\times Lenght^2\times Time^{-1}$,
so that \ref{eqq} has introduced a "mass" into the system.
Applying \ref{eqq} to the constraint gives a modified Klein-Gordon equation
\be
\la\ps=-\hb^2\left(\bx+\fr{16\te^2}{\hb^2U^6}\right)\ps=0.
\lb{eqk}
\ee
Defining
\be
S\equiv-i\hb\ln\ps
\lb{eqsd}
\ee
the modified Klein-Gordon equation \ref{eqk} becomes
\be
-i\hb S^s_{.a}+S^aS_a-\fr{16\te^2}{U^6}=0,
\lb{eqsa}
\ee
expanding in terms of $\hb$ using
\be
S_a=\Pi_a+\hb\ep_a+\Or(\hb^2),
\lb{eqex}
\ee
the $\hb^0$ term is just the constraint \ref{eqco},
the $\hb^1$ term is
\be
-\imath\Pi^a_{.a}+2\ep_a\Pi^a=0,
\lb{eqep}
\ee
For $\te=0$,  the lagrangians $l_1$ and $l_2$ vanish as does $\Pi$,
so that to lowest order $\hb^0$, $S_a=0$,  implying that the wavefunction $\ps$
is a constant to lowest order,  thus for $\te=0$ the wavefunction has no dynamical information
corresponding to the classical theory.

$U$ remains unchanged during quantization,  
but once a solution $\ep$ to \ref{eqex} is known,
one would hope to be able to calculates the $\hb^1$ order correction to $U$ and hence $g$.
There is a problem with trying this,
as $\Pi$ is only partially invertible \ref{eqp1} \& \ref{eqp2}
this cannot be done without an additional assumption.
Here this assumption is that $\te$ remains negligible to order $\hb^1$
in the quantum theory,  
then it is possible to find the correction to $U$ from \ref{eqp2},
denoting the quantum quantities with a "*" $U$ becomes
\be
U^*_a=(4\te)^\fr{1}{3}S^{_\fr{4}{3}}S_a.
\lb{ust}
\ee
Substituting for $S$ using \ref{eqex} and expanding
\be
U^*_a=U_a+\fr{\hb}{4}\fr{U^4}{\te}\left(\ep_a-\fr{8}{3}\fr{U_c\ep^c}{U^2}U_a\right)+\Or(\hb^2).
\lb{ueu}
\ee
It is now possible to investigate whether the assumption that $\te$ is negligible by noting
\ber
&-\te^*\equiv U^{*a}_{~.;a}-U^{*-2}\left(U^{*2}\right)^\circ=-\te\n
&+\fr{\hb}{4}\left[\fr{U^4}{\te}\left(\ep^a-\fr{8}{3}U^{-2}U^c\ep_cU^a\right)\right]_a
+\fr{5\hb}{6}\left[U^{-2}\left(\fr{U^4\ep_aU^a}{\te}\right)^\circ
                   -\fr{(U^2)^\circ\ep_aU^a}{\te}\right]+\Or(\hb^2).
\lb{tsu}
\ear
Substituting for $U^*$ the change in the metric is
\be
g^*_{ab}=g_{ab}+\fr{\hb}{\te}\left(U_c\ep^cU_aU_b-U^2U_{(a}\ep_{b)}\right)+\Or(\hb^2).
\lb{gsu}
\ee
The change in the metric can also be directly calculated from the wavefunction
\be
g^*_{ab}-g_{ab}-2U^{-2}U_aU_b
=-2U^{-2}U^*_aU^*_b
=-2S^{-\fr{4}{3}}S_aS_b
=2\hb^2(-i\hb\ln\ps)^{-\fr{4}{3}}\ps^{-2}\ps_a\ps_b.
\lb{gss}
\ee

The modified Klein-Gordon equation \ref{eqk} can be studied for particular examples,
for example in Robertson-Walker spacetime \ref{pdrw} it is 
\be
\ps_{00}+3\fr{R_0}{R}\ps_0+\fr{144}{\hb^2}\fr{R^2_0}{R^2}\ps-\fr{l(l+2)}{R^2}\ps=0,
\lb{kmrw}
\ee
where the last term comes from decomposing the "spatial" part into spherical harmonics
\cite{mdr32}\S4.1.   For the Milne universe,  which is flat when $k=-1$ \cite{mdr20},
$R=t$ and \ref{kmrw} has solution
\be
\ps=At^{-1\pm\sq{1-\al}},~~~~~~~
\al=\fr{144}{\hb^2}-l(l+2),
\lb{mups}
\ee
so that $g^*$ is of the form $f\hb^2/t^2$.   For deSitter space \cite{HE}p.125,
$R=\exp(\sq{\La/3}t)$ and when $l=0$ \ref{kmrw} has solution
\be
\ps=A\exp\left(\fr{1}{2}\sq{3\La}(-1\pm\sq{1-16/\hb^2})t\right),
\lb{dsps}
\ee
so that $g^*$ is of the form $f\hb^2$.
\section{Conclusion.}\lb{conc}
The transformation between some specific positive definite spaces and Lorentzian
spacetime can be achieved via a line element field \ref{pdsc} \ref{pdrw}.
This can be generalized to shift-free and then arbitrary metrics;
there is a problem of what the contravariant form of the metric should be,
consistency requires \ref{n2}.
Once the Lorenztian metric has been expressed in terms of a positive definite
metric and a vector field it is possible to study first derivatives.
In \ref{eq6} $L=0$ so that the Lorentzian connection splits up into the positive
definite connection and a contorsion term constructed from the line element field $U$;
this is similar to many other systems,  such as those involving Schouten \cite{schouten}
geometries and weak perturbations \cite{mdr32};
that $L=0$ perhaps is not surprising as the decomposition 
of the Lorentzian metric is covariant.
The form of the contorsion tensor \ref{eq7} involves a lot of terms when expressed 
soley in terms of $U$,  however using rotation and shear it takes a simler form \ref{n9}.
Covariant derivatives in Lorentzian spacetime and the positive definite space are equated
via \ref{6b},  so that the difference is expressible as $V_cK^c_{ab}$ 
and this is proportional to the expansion of $U$,
changing spaces has the effect of changing the sign of the expansion.
Second derivatives of the line element field $U$ can be calculated
once the contorsion $K$ is known via \ref{eq10}.

To quantize the system it is necessary to have more information,
such as what the Lagrangian and momentum are.
Here the vacuum-Einstein-Hilbert Lagrangian is assumed \ref{eq11},
and further that it can be decomposed into a positive definite part
and a line element field part which have well-defined and usefull variations.
Variations with respect to the metric and the line element field \ref{12b} can be done,
however of more use is variation with respect to $\dot{U}$ 
which is taken to give a momentum \ref{eqm};
variations with respect to dotted quantities also occur in the quantization of
perfect fluids \cite{mdr27}.
The momentum obeys the constraint \ref{eqco}.
The two-sided nature of $U$,
the Lorentzian metric is invariant under $U\rightarrow-U$;
and the ability to use $U$ of different sizes to construct the Lorentzian metric
do not seem to lead to further constraints.
Quantization can be achieved via \ref{eqq}.
The problem with this is that it introduces a mass into the system.
The classical theory is just a theory involving lenght and time,
however Planck's constant has dimensions $Mass\times Length^2\times Time^{-1}$,
so that using it in quantization introduces new quantities of dimension $Mass$.
Theories,  such as the vacuum-Einstein equations,  involving just lenght and time
are usually reversible,  in the sense that the sign on the time coordinate can be 
changed and the field equations still obeyed;
however this is no longer necessarily the case once quantities of dimensions
of mass have been introduced,  
as illustrated by the fact that things fall down not up.
This is not only a problem for the theory under study here,
similiarly using $\hb$ in quantization of the vacuum-Einstein equations will
introduce a mass.
The specific wavefunctions \ref{mups} and \ref{dsps} illustrate the above,
of the two terms in the square root one is dimensionless "$1$" and the other
is dimensionfull and proportional to $\hb^{-2}$.
A way of avoiding the above is to divide $\hb$ by the Planck mass 
or perhaps an arbitrary mass so that objects of dimensions of mass 
no longer occur in the quantum system;
also by analogy with the point particle one could perhaps pre-multiply
the line element field Lagrangian by an arbitrary $m$,
but on the analysis so far such an $m$
does not occur naturally,
perhaps it might do so in an extended theory which in some way incorporates
that $U$ is not necessarily of unit size.
Any given Lorentzian metric can be constructed from many different sets of a 
positive definite metric and a line element field.
For example flat spacetime can be expressed by the Minkowski metric and this
can be constructed from a diagonal metric and unit expansion free line element field;
also flat spacetime can be expressed by the Milne universe \ref{mups}
for which $U$ has expansion.
In the first case there is no expansion and hence no momentum or quantum theory,
in the second there is with wavefunction \ref{mups}.
Thus it might be that "Euclidean" quantum gravity expresses the full quantum nature
of a Lorentzian spacetime if the relating line element field is expansion free;
however the main application of such theories is to the early universe where
expansion is the most salient feature.  

\end{document}